\documentclass[twocolumn,resetfootnote]{aastex701}

\usepackage{amssymb,amsfonts,amsmath,mathtools,needspace,enumitem,etoolbox,
graphicx,physics,microtype,afterpage,gensymb,tabularx,bigints,relsize,soul,xspace,xcolor}
\usepackage{url}
\usepackage{circuitikz}

\newcommand{\milan}{Dipartimento di Fisica ``G. Occhialini'', 
Universit\`a degli Studi di Milano-Bicocca, Piazza della Scienza 3, 20126 Milano, Italy}
\newcommand{\infn}{INFN, Sezione di Milano-Bicocca, 
Piazza della Scienza 3, 20126 Milano, Italy}

\newcommand{\rate}{\mathrm{d}_{\theta}N}

\newcommand{\tri}{\Delta}
\newcommand{\Tobs}{T_{\mathrm{obs}}}
\newcommand{\many}[1]{\left\{ #1 \right\}}
\newcommand{\Msun}{\mathrm{M}_{\odot}}
\newcommand{\Mtot}{M_{\mathrm{tot}}}

\newcommand{\Ndet}{N_{\mathrm{det}}}

\newcommand{\Nv}{N_{\mathrm{v}}}
\newcommand{\Nc}{N_{\mathrm{c}}}
\newcommand{\Np}{N_{\mathrm{par}}}
\newcommand{\R}{\mathcal{R}}
\newcommand{\volrate}{\mathrm{d}_{\theta} \R}

\begin{document}

\title{\large
Where did heavy binaries go?\\[3pt]
Gravitational-wave populations using Delaunay triangulation with optimized complexity
}

\author[orcid=0000-0002-3582-2587,sname=Tenorio]{Rodrigo Tenorio}
\affiliation{\milan}
\affiliation{\infn}
\email[show]{\href{mailto:rodrigo.tenorio@unimib.it}{rodrigo.tenorio@unimib.it}}

\author[orcid=0000-0002-2685-1538,sname=Toubiana]{Alexandre Toubiana}
\affiliation{\milan}
\affiliation{\infn}
\email{alexandre.toubiana@unimib.it}

\author[orcid=0000-0002-1789-7876,sname=Bruel]{Tristan Bruel}
\affiliation{\milan}
\affiliation{\infn}
\email{tristan.bruel@unimib.it}

\author[orcid=0000-0002-0933-3579,sname=Gerosa]{Davide Gerosa}
\affiliation{\milan}
\affiliation{\infn}
\email{davide.gerosa@unimib.it}

\author[orcid=0000-0002-1671-3668,sname=Gair]{Jonathan R. Gair}
\affiliation{Max Planck Institute for Gravitational Physics (Albert Einstein Institute), Am Mühlenberg 1, 14476, Potsdam, Germany}
\email{jonathan.gair@aei.mpg.de}

\begin{abstract}
We investigate the joint mass-redshift evolution of the binary black-hole merger rate in the latest
Gravitational-Wave Transient Catalog, GWTC-4.0. We present and apply a novel non-parametric
framework for modeling multi-dimensional, correlated distributions based on Delaunay triangulation.
Crucially, the complexity of the model---namely, 
the number, positions, and weights of triangulation nodes---is inferred directly
from the data, resulting in a highly efficient approach that requires about 
one to two orders of magnitude fewer parameters and significantly less calibration 
than current state-of-the-art methods. 
We find no evidence for a peak at $\Mtot \sim 70\,\Msun$ at low redshifts ($z \sim 0.2$), 
where it would correspond to the $m_1 \sim 35\,\Msun$ feature reported in redshift-independent 
mass spectrum analyses, and we infer an increased merger rate at high redshifts 
($z \sim 1$) around those masses, compatible with such a peak.
When related to the time-delay distribution from progenitor formation to binary black-hole merger,
our results suggest that sources contributing to the $m_1 \sim 35\,\Msun$ feature
follow a steeper (shallower) time-delay distribution at high (low) redshifts. 
This hints at contributions from different formation channels---for example 
dense environments and isolated binary evolution, respectively---although
firm identification of specific formation pathways will require further observations and analyses.
\end{abstract}


\section{Introduction}

The increasing number of binary black holes (BBHs) observed by gravitational-wave (GW) 
interferometers~\citep{LIGOScientific:2018mvr,LIGOScientific:2020ibl,KAGRA:2021vkt,
LIGOScientific:2021usb,LIGOScientific:2025slb} deepens our understanding of their
population in the Universe~\citep{LIGOScientific:2025pvj}, 
eventually revealing their origin out of the multiple proposed formation 
channels~\citep{Mandel:2018hfr,Mapelli:2021taw}. A crucial step in this direction is the identification of correlations between parameters and the presence of distinct subpopulations (e.g. \citealt{Callister:2024cdx} and references therein)

A key question is whether, and how, the BBH population evolves with redshift and, crucially, whether this evolution depends on the BBH properties.
BBHs are the end products of stellar evolution; 
therefore, their population should depend on the star-formation history and properties 
of galaxies such as metallicity~\citep{Mapelli:2019bnp,Neijssel:2019irh,vanSon:2021zpk,
deSa:2024otm,Broekgaarden:2021iew}.
Moreover, BBHs involving remnants of previous mergers, 
formed through dynamical encounters,
also contribute to the evolution of the population's properties~\citep{Gerosa:2021mno,Ye:2024ypm,Torniamenti:2024uxl}. 
Finally, the redshift evolution encodes the relative contribution of the 
different BBH formation 
channels throughout cosmic history~\citep{Zevin:2020gbd,Mapelli:2021gyv,Sedda:2021vjh}.

Evidence for an evolution of the effective spin distribution with redshift~\citep{Biscoveanu:2022qac}
was found on the third gravitational-wave transient catalog (GWTC-3) by LIGO, Virgo, and KAGRA, 
and was further strengthened in their fourth catalog (GWTC-4.0;~\citealt{LIGOScientific:2025pvj}). 
In contrast, several analyses of GWTC-3 found no evidence for or against a redshift evolution in 
the mass distribution~\citep{Ray:2023upk,Heinzel:2024hva,Sadiq:2025aog,Lalleman:2025xcs,Gennari:2025nho}, 
and first analyses on GWTC-4.0 find similar conclusions~\citep{LIGOScientific:2025pvj}.\footnote{
While \cite{Rinaldi:2023bbd} did report evidence for such an evolution,
their treatment of selection effects is not accurate~\citep{Essick:2023upv,Toubiana:2025syw}.
}

In this work, we investigate the joint mass-redshift evolution of the BBH merger rate 
using a novel non-parametric multi-dimensional approach. 
In particular, we reconstruct the joint mass–redshift BBH merger rate using
Delaunay triangulation~\citep{Delaunay_1934aa} and barycentric interpolation. 
Both the number and locations of the triangulation vertices, as well as their associated weights, 
are inferred directly from the data through trans-dimensional Bayesian inference~\citep{Toubiana:2023egi}. 
The use of a data-driven interpolation scheme assumes no specific functional (in)dependence
between mass and redshift; this allows us to probe a broader parameter space compared 
to models with closed-form correlations.

Using GWTC-4.0 data\footnote{
We analyze the 153 BBH events considered by~\citet{LIGOScientific:2025pvj}, see their Sec. 6.
Note this includes only those events with less than 1\% posterior support 
for component masses below~$3 \, \Msun$.
}, 
we report a distinct difference in the distribution of masses at $z=0.2$ 
and $z=1.0$ at $\Mtot \sim 70 \, \Msun$,
with no evidence for the presence of a peak in the merger rate at said masses for $z = 0.2$.
We then discuss the astrophysical implications of our findings.

\section{Mass-redshift correlation}

\begin{figure}
\centering
    \resizebox{1\columnwidth}{!}{%
        \begin{circuitikz}
\tikzstyle{every node}=[font=\LARGE]
\fill[cyan!20] (9.75,16.75) -- (15.5,16) -- (14,19) -- cycle;
\draw [ color={rgb,255:red,154; green,153; blue,150}, , line width=0.8pt](9.75,16.75) to[short] (9.75,16.75);
\draw [line width=0.8pt, short] (9.75,16.75) -- (15.5,16);
\draw [line width=0.8pt, short] (14,19) -- (15.5,16);
\draw [line width=0.8pt, short] (14,19) -- (9.75,16.75);
\draw [ color={rgb,255:red,237; green,51; blue,59}, line width=1pt, dashed] (9.75,16.75) -- (13.25,17.5);
\draw [ color={rgb,255:red,237; green,51; blue,59}, line width=1pt, dashed] (14,19) -- (13.25,17.5);
\draw [ color={rgb,255:red,237; green,51; blue,59}, line width=1pt, dashed] (15.5,16) -- (13.25,17.5);
\node at (13.25,17.5) [circ, color={rgb,255:red,237; green,51; blue,59}] {};
\node [font=\large, color={rgb,255:red,237; green,51; blue,59}] at (13.25,18) {$\theta$};
\node [font=\normalsize] at (14,17.5) {$b_1(\theta)$};
\node [font=\normalsize] at (13,16.75) {$b_2(\theta)$};
\node [font=\normalsize] at (12.5,17.75) {$b_3(\theta)$};
\node [font=\normalsize] at (9.1, 16.75) {$(v_1, w_1)$};
\node [font=\normalsize] at (14, 19.5) {$(v_2, w_2)$};
\node [font=\normalsize] at (16.3, 16.1) {$(v_3, w_3)$};
\node [font=\normalsize] at (11.25,18.25) {$S(\theta)$};
\draw [ color={rgb,255:red,154; green,153; blue,150}, , line width=1pt](8.5,19.75) to[short] (17.75,19.75);
\draw [ color={rgb,255:red,154; green,153; blue,150}, , line width=1pt](8.5,19.75) to[short] (8.5,14.5);
\draw [ color={rgb,255:red,154; green,153; blue,150}, , line width=1pt](8.5,14.5) to[short] (17.75,14.5);
\draw [ color={rgb,255:red,154; green,153; blue,150}, , line width=1pt](17.75,19.75) to[short] (17.75,14.5);
\draw [ color={rgb,255:red,154; green,153; blue,150}, line width=1pt, short] (8.5,14.5) -- (9.75,16.75);
\draw [ color={rgb,255:red,154; green,153; blue,150}, line width=1pt, short] (15.5,16) -- (8.5,14.5);
\draw [ color={rgb,255:red,154; green,153; blue,150}, line width=1pt, short] (15.5,16) -- (17.75,14.5);
\draw [ color={rgb,255:red,154; green,153; blue,150}, line width=1pt, short] (15.5,16) -- (17.75,19.75);
\draw [ color={rgb,255:red,154; green,153; blue,150}, line width=1pt, short] (14,19) -- (17.75,19.75);
\draw [ color={rgb,255:red,154; green,153; blue,150}, line width=1pt, short] (8.5,19.75) -- (14,19);
\draw [ color={rgb,255:red,154; green,153; blue,150}, line width=1pt, short] (8.5,19.75) -- (9.75,16.75);
\node [font=\large, fill=white, fill opacity=1, text opacity=1, inner sep=2pt, draw={rgb,255:red,154; green,153; blue,150}, thin] at (8.5, 19.75) {$W_1$};
\node [font=\large, fill=white, fill opacity=1, text opacity=1, inner sep=2pt, draw={rgb,255:red,154; green,153; blue,150}, thin] at (8.5, 14.5) {$W_2$};
\node [font=\large, fill=white, fill opacity=1, text opacity=1, inner sep=2pt, draw={rgb,255:red,154; green,153; blue,150}, thin] at (17.75, 19.75) {$W_3$};
\node [font=\large, fill=white, fill opacity=1, text opacity=1, inner sep=2pt, draw={rgb,255:red,154; green,153; blue,150}, thin] at (17.75, 14.5) {$W_4$};
\node at (9.75,16.75) [circ, color={rgb,255:red,237; green,51; blue,59}] {};
\node at (15.5,16) [circ, color={rgb,255:red,237; green,51; blue,59}] {};
\node at (14,19) [circ, color={rgb,255:red,237; green,51; blue,59}] {};

\end{circuitikz}
    }
    \caption{
        Delaunay triangulation to model the differential merger rate 
        $\log_{10}\rate$ across two variables (on the horizontal and vertical axes, respectively).
        The central red dot represents a location $\theta$ where the rate needs to be computed.
        The highlighted area $S(\theta)$ represents the
        triangle (\emph{simplex} in higher dimensions) containing $\theta$ whose vertices $v_i$ and
        weights $w_i$ are inferred from the data.
        The position of the four corners is fixed in advance, and their weights $W_{i}$ 
        are inferred from the data.
        The rate $\log_{10}\rate$ is computed by interpolating the weights at the vertices of $S(\theta)$ 
        using the barycentric coordinates $b_i(\theta)$ associated to $\theta$.
        }
    \label{fig:tri_tikz}
\end{figure}
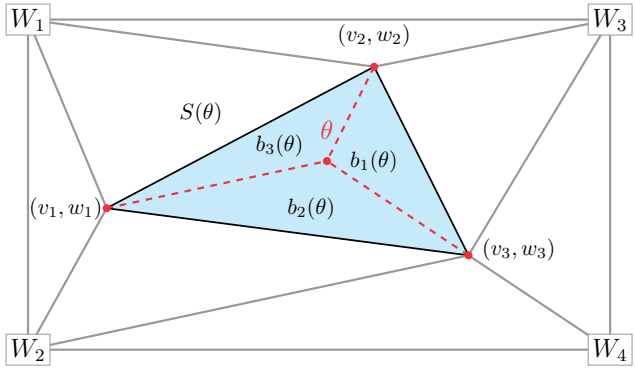

We denote the targeted set of source parameters by $\theta$, consisting of total mass 
$\Mtot$, mass ratio $q \leq 1$, redshift $z$, spin $\chi_{1,2}$ and  cosine of the spin tilts $\cos \vartheta_{1,2}$. To capture potential dependencies between 
$\Mtot$  and $z$, we model the differential number of events $\rate(\theta | \Lambda)$ as  
\begin{equation}
    \begin{aligned}
    &\log \rate(\theta| \Lambda) = 
    \tri(\Mtot,z|\Lambda_{\tri}) + \log\big[ p(q|\Mtot,\Lambda_{q}) \big]
    \\ 
    &+\log \big [p(\chi_1|\Lambda_{\chi}) p(\chi_2|\Lambda_{\chi})\big] +\log \big[p(\cos\vartheta_1,\cos\vartheta_2|\Lambda_{\vartheta})
        \big] \,, \label{eq:pop_model}
    \end{aligned}
\end{equation}
where $\tri(\Mtot,z|\Lambda_{\tri})$ represents the (natural log) differential rate as computed by 
barycentric interpolation using Delaunay triangulation.
The hyperparameters $\Lambda_{\tri}$ are the number of triangulation vertices, 
their locations, and the value of the differential rate at those vertices.
Figure~\ref{fig:tri_tikz} gives a schematic description of our model; 
further details are provided in Appendix~\ref{app:delaunay}. 
We model the mass ratio distribution as a broken power-law, 
which is compatible with the marginalised $q$ distribution found 
in~\citet[][see Appendix~\ref{app:q} for details]{LIGOScientific:2025pvj}.
For the spin magnitudes and tilt angles we assume the same functional forms as 
in the default model  by~\cite{LIGOScientific:2025pvj}. 
The hyperparameters $\Lambda=(\Lambda_{\tri},\Lambda_q,\Lambda_{\chi},\Lambda_{\vartheta})$ are inferred through hierarchical Bayesian inference (see Appendix~\ref{sec:selection} for details). 
The differential volumetric rate of GW events is then given by
\begin{equation}
    \volrate = \left( \frac{\Tobs}{1 + z} \frac{\mathrm{d} V_{\mathrm{c}}}{\mathrm{d}z} \right)^{-1} \rate \,,
    \label{eq:vol_rate}
\end{equation}
where $\Tobs$ is the observation time. For completeness, we denote as $m_{1,2}$ the primary and secondary masses. 

We limit the domain of inference for the $(\Mtot, z)$ distribution to 
$[6,  350] \, \Msun \times [0, 2.5]$.
Priors on triangulation vertices  are uniform within this domain,
and weights are uniformly distributed along $(-20, 15)$.
The number of triangulation nodes is allowed to vary between 4 and 100. 
Inference is conducted using reversible-jump Markov chain Monte Carlo as 
implemented in \texttt{eryn}~\citep{Karnesis:2023ras}. We have verified 
our findings to be robust against the choice of prior 
(tests are reported in Appendices~\ref{app:q} and~\ref{sec:selection}).

\begin{figure*}
    \centering
    \includegraphics[width=\columnwidth]{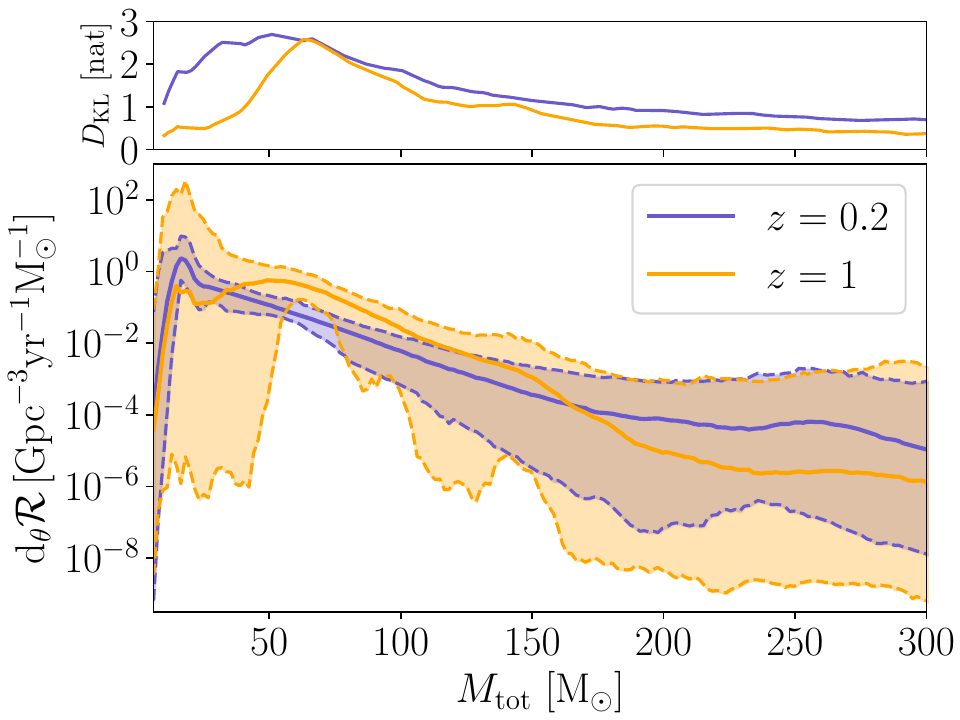}
    \includegraphics[width=\columnwidth]{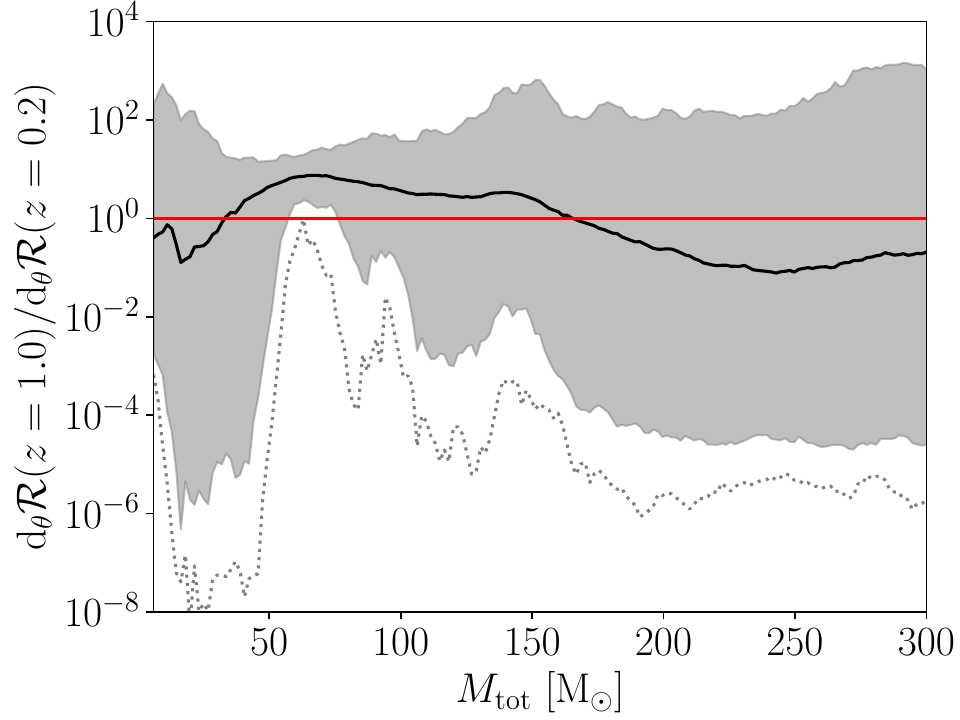}
    \caption{
        \emph{Left:} 
        Posterior distribution of the volumetric differential rate reconstructed using Delaunay 
        triangulation for two representative redshift values, $z=0.2$ (purple) and $z=1$ (orange).
        Solid curves indicate the medians while shaded regions 
        encompass 90\% symmetric credible intervals.
        The upper panel shows the Kullback-Leibler divergence between the
        posterior and prior volumetric differential rate distributions.
        \emph{Right:}
        Posterior distribution of the ratio of the volumetric differential rate 
        at the same two representative redshift values. 
        The black solid curve denotes the posterior median. The shaded region contains
        the symmetric 90\% credible interval.
        The horizontal red line corresponds to $\volrate(z=0.2)=\volrate(z=1)$. 
        The dotted line corresponds to the $1.5\%$ credibility interval.
        The observed difference in rates is thus inconsistent with no evolution of the rate with $z$ at 
        $\sim 98\%$ credibility, but remains consistent with a mass-uniform evolution
        (the envelope is consistent with a straight line).
    }
    \label{fig:Mtot_z}
\end{figure*}

The left panel of Fig.~\ref{fig:Mtot_z} shows the posterior distribution of the 
volumetric differential rate at two representative redshift values ($z=0.2$ and $z=1.0$).
Their behavior is different:
\begin{itemize}
\item At $z=0.2$,
the merger rate shows a distinctive peak at $\Mtot \sim 20\, \Msun$ and decays in 
a featureless, power-law-like manner. 
\item For $z=1$, the merger rate
increases with respect to that at low redshifts for \mbox{$\Mtot \sim  70 \,\Msun$} and appears to have an excess of systems at those masses.
\end{itemize}
The right panel of Fig.~\ref{fig:Mtot_z} shows the credibility for the 
$z\sim 1$, $\Mtot \sim 70\, \Msun$ feature, which we quantify as the posterior probability
of \mbox{$\volrate(\Mtot, z = 1) > \volrate( \Mtot, z = 0.2)$}. This value is $\gtrsim 95\%$ 
for $\Mtot \sim 70\,\Msun$, and reaches a maximum of $98\%$ at $\Mtot \sim 63 \,\Msun$.
Said increase with redshift is compatible with that found by~\citet{LIGOScientific:2025slb},
where the merger rate is modeled as a function of redshift (but not mass) with an ansatz $\volrate \propto (1+z)^\kappa$ and returned $\kappa \sim 3$.

While we cannot definitively claim that the mass distribution at $z=1.0$ exhibits a peak at 
$\sim 70\,\Msun$ owing to the error bars, a corresponding feature at $m_1 \sim 35 \,\Msun$ 
has been consistently identified since GWTC-2 \citep{LIGOScientific:2020kqk}.
Our findings suggest that this feature is associated with high-redshift BBHs 
and disappears by $z = 0.2$. 
In Appendix~\ref{app:m1_z}, we show the results of applying our method to $(m_1,z)$ 
and illustrate how our discussion of the $70 \,\Msun$ feature similarly applies to 
the $35 \,\Msun$ peak in $m_1$. Altogether, these results indicate that this high-mass feature 
in the BBH mass distribution~\citep{LIGOScientific:2020kqk,KAGRA:2021duu,LIGOScientific:2025pvj}
varies with redshift. These results are also compatible with~\citet{Rinaldi:2025emt}, 
who reported tentative evidence for a steeper redshift evolution associated to the $m_1 \sim 35 \, \Msun$ peak 
than that observed for the power-law component in the GWTC-3 catalog.
Future analyses should consider targeting such behavior with more strongly modeled approaches.

\begin{figure}
    \centering
    $\,$\\[5pt]
    \includegraphics[width=\columnwidth]{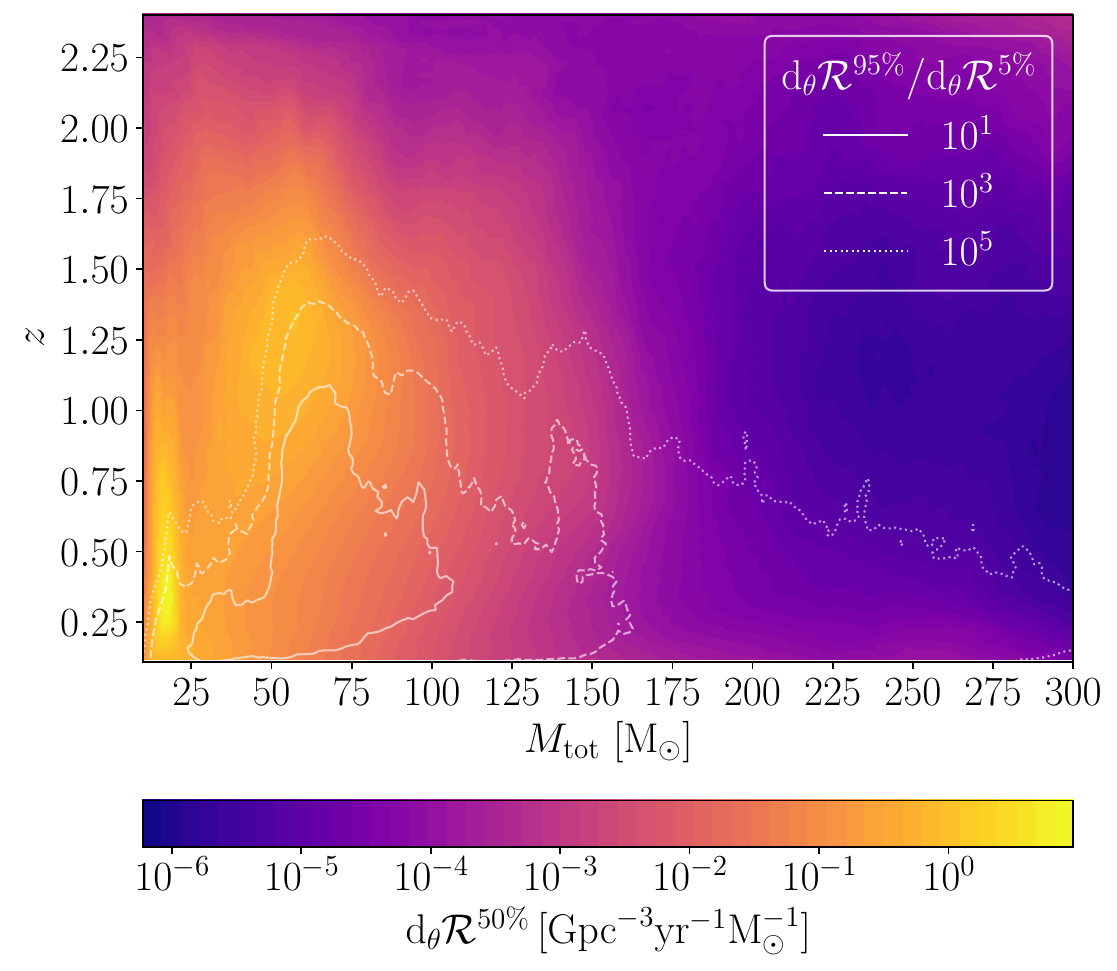}
    \caption{
        The color scale shows the median (50\% quantile) posterior volumetric rate 
        inferred using Delaunay triangulation on GWTC-4.0. 
        The white contours indicate the relative uncertainty on the rate itself,
        defined as the ratio between the 95\% quantile and the 5\% quantile.
        }
    \label{fig:KL_2d}
\end{figure}

Figure \ref{fig:KL_2d} presents the outcome of our inference across the $(\Mtot, z)$ parameter space.
The region of high uncertainty roughly aligns with the detection horizon beyond
which no BBH mergers were detected;
conversely, low uncertainties align with data-dominated regions.
We find that BBHs with different masses merge more prominently at different redshifts 
and that the high-mass prominence appears around  \mbox{$z \sim 0.7$}. 
However, given current uncertainties, no definitive statement about distinct BBH sub-populations 
can yet be made.

Our findings are not sufficient to establish whether the mass distribution changes with redshift,
as the relative difference in rates is consistent with a mass-independent rescaling 
(see right panel of Fig.~\ref{fig:Mtot_z}). 
The key result of this letter is the disappearance of the high-mass peak at low redshift. 

We characterize the high-redshift appearance of the \mbox{$\sim 70\,\Msun$} feature 
by computing Kullback-Leibler divergence ($D_{\mathrm{KL}}$; \citealt{kl_div})
between the posterior and the prior on the rate induced by our prior on $\Lambda_{\tri}$
For reference, $D_{\mathrm{KL}} \lesssim 0.1\,\mathrm{nat}$ between two independent prior draws.
As show in the upper left panel of Fig.~\ref{fig:Mtot_z},
we find significantly larger values $D_{\mathrm{KL}}\sim 2.5 \, \mathrm{nat}$ around $\Mtot \sim 70 \,\Msun$,
showing that these results are primarily driven by the observed data.
High $D_{\mathrm{KL}}$ values align with low relative uncertainties
in Fig.~\ref{fig:KL_2d} ($\volrate^{95\%} / \volrate^{5\%}\lesssim 10$), 
where the inference is dominated by the observed events.

Our results differ from those by \cite{LIGOScientific:2025pvj},
which reported no evidence for a mass-dependent redshift evolution of the merger
rate using both copulas~\citep{Adamcewicz:2022hce} and binned Gaussian processes~\citep{Ray:2023upk}. 
Copulas are suited to identify linear correlations, but not the kind of multi-modal structures
inferred in Fig.~\ref{fig:KL_2d}. In the binned Gaussian process approach, 
the high-mass feature appears less pronounced at lower redshift, consistent with our findings,
though large error bars blur the trend. 
Their results were shown as probability density functions of $m_1$ across redshift bins,
which are normalized by the total number of events, a very uncertain quantity.
As a result, error bars end up broadening.
This issue is avoided when working with the volumetric merger rate.
The counterpart limitation is that the nonoverlap of confidence bands of the volumetric differential rate at different redshifts is not, by itself, sufficient to establish redshift evolution of the mass distribution.
This is shown in the right panel of Fig.~\ref{fig:Mtot_z}, where a straight line remains consistent 
with the 90\% credibility interval of the rate ratio, which is thus compatible with an overall, 
rigid rescaling of the $\volrate$. 
Finally, unlike the binned method, our interpolation points are data-driven rather than fixed, 
potentially capturing finer features by construction.

We also infer a peak in the merger rate at \mbox{$\Mtot \sim 20 \, \Msun$} for $z=0.2$, 
but the lack of detections at higher redshifts prevents us from confirming or disproving 
its existence for $z \gtrsim 0.2$.  
At the high-mass end, the bump at $z = 0.2$ around $\Mtot \sim 250 \,\Msun$ is likely driven by a single event, 
GW231123~\citep{LIGOScientific:2025rsn}, while the the apparent increase in the rate for $\Mtot \sim 150\,\Msun$ 
at $z = 1$ aligns with the inferred source properties of GW190521~\citep{LIGOScientific:2020iuh}.
We caution against overinterpreting such features, which indeed correspond to low values
of $D_{\rm KL}\sim 1\,{\rm nat}$.

Our inference on the mass-ratio distribution is consistent with a larger density
at $q \gtrsim 0.6$, similarly to what was reported by~\cite{LIGOScientific:2025pvj}.
Details are reported in Appendix~\ref{app:q}.

The triangulation scheme that best fits the $(\Mtot,z)$ portion of the population 
has ${17}_{-11}^{+25}$ nodes, corresponding to $56^{+75}_{-33}$ free parameters (90\% credible interval).
This is one to two orders of magnitude fewer than those needed by~\cite{Heinzel:2024hva},
and requires less tuning compared to~\cite{Ray:2023upk}, which computes the 
final number of bins by re-running the analysis using increasingly finer grids.
Crucially, both of those approaches rely on a fixed two-dimensional grid and
are therefore expected to scale poorly with the number of dimensions,
whereas Delaunay triangulation is naturally applicable to higher-dimensional correlations.
The method proposed by \citet{Payne:2022xan} and \cite{Guttman:2025jkv} 
requires a comparable number of parameters, 
but returns the maximum likelihood histogram of data without an associated error.
The approach of~\cite{Sadiq:2021fin},
while having a very small number of free parameters,
seems to suffer from edge effects where data is scarce or absent;
our method correctly returns the prior in these regions of parameter space.

\section{Astrophysical interpretation}\label{sec:astro}

\begin{figure}
    \centering
    \includegraphics[width=\columnwidth]{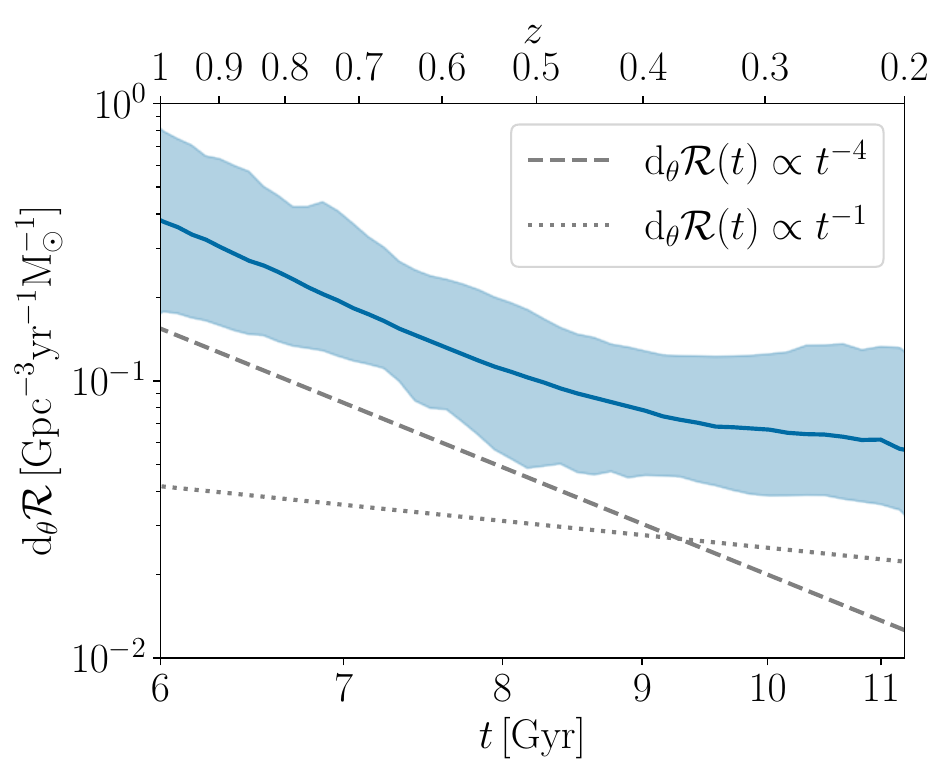}
    \caption{
        Evolution of the volumetric differential merger rate as a function of cosmic time 
        $t$ in log-log scale evaluated at $M_\mathrm{tot}=63\,\Msun$, 
        i.e. a nominal location where evolution with redshift is  tightly constrained. 
        The solid curve indicates the median while the shaded region encloses the 90\%
        symmetric credible interval.
        The dashed (dotted) line corresponds to a merger rate evolving 
        as a power law with index $-1$ $(-4)$  and an arbitrary normalization for reference.
        }
    \label{fig:td_distro}
\end{figure}

Our triangulation-based reconstruction suggests that the known
feature in the BBH mass spectrum at  \mbox{$\Mtot\sim 70 \,\Msun$ \, ($m_1 \sim35\,\Msun$)}
is primarily due to mergers at \mbox{$z\gtrsim 0.7$} and fades by $z \sim 0.2$.
More specifically, as shown in Fig.~\ref{fig:td_distro}, 
the evolution of $\volrate$ as a function of cosmic time $t$ between $z=1$ and $z=0.2$
appears to follow two regimes, with $\volrate(t) \propto t^{-1}$ at low 
redshift and a much steeper evolution closer to $\volrate(t) \propto t^{-4}$ at higher redshift.
We stress these are indicative trends and not rigorous fits, which would require more data.

Population-synthesis models make predictions for the dependence of the merger rate on the delay time
$t_{\mathrm{delay}}$, i.e. the time between progenitor formation and BBH merger.
In the, likely simplistic, assumption where all stellar progenitors of merging BBHs form in a burst at cosmic time $t_0$,
the merger rate traces the delay-time distribution, 
$\volrate(t) \sim \mathrm{d} N / \mathrm{d} t_{\mathrm{delay}}$ with
\mbox{$t_{\mathrm{delay}} = t - t_0$}.
The results in Fig.~\ref{fig:td_distro} are thus informative about the astrophysical channels 
that generate the $\Mtot\sim 70 \,\Msun$ feature.

At low redshift,  \mbox{$\volrate(t) \propto t^{-1}$} 
is compatible with typical predictions for isolated binary evolution, including sources ejected from stellar clusters
~\citep[see e.g.][]{Dominik:2012kk,Belczynski:2016obo,Eldridge:2016ymr,Lamberts:2016txh,Mapelli:2017hqk}.

By contrast, the steeper behavior at higher redshift $\volrate(t) \propto t^{-4}$ 
is in line with numerical simulations of cluster evolution where BBHs merge inside their host 
clusters~\citep[e.g.][]{Rodriguez:2018rmd,Sedda:2023qlx}.
Moreover, these dynamically assembled BBHs are, on average, 
more massive than those from isolated evolution
~\citep{Rodriguez:2016kxx,DiCarlo:2019pmf, 
Antonini:2020xnd,Sedda:2023qlx}, which would explain their prominence at
\mbox{$\Mtot\sim 70 \,\Msun$ \, ($m_1 \sim35\,\Msun$)}~\citep{Antonini:2022vib,Bruel:2025sdq}.

In short, a possible explanation of our findings is that the dense‑environment channels, 
which are predicted to form BBH with higher $z$ and shorter delay, 
are responsible for the $\Mtot \sim 70 \, \Msun$ feature, 
with isolated (and ejected) binaries providing the longer‑delay background 
that dominates at low redshift.

Massive, dense globular clusters match the requirements for this explanation, 
as they likely tend to form at earlier  cosmic times with respect to the global formation 
of stars in the  Universe~\citep{2018MNRAS.480.2343C,2019MNRAS.482.4528E} and thus naturally 
give rise to a  high-mass contribution to the BBH merger rate that switches off at low redshifts.
Numerical analyses of this scenario predict a high-mass BBH peak at  $z \gtrsim2$ that extends 
to lower redshifts~\citep{Ye:2024ypm,Torniamenti:2024uxl},
suggesting that our observation could in fact correspond to the lower-redshift tail of this peak. 
Similarly, AGN disks are another dense environment favoring hierarchical mergers which can also
produce heavy BBHs with short delays~\citep{Yang:2019cbr,Tagawa:2019osr,Santini:2023ukl}.

Pulsational pair-instability supernovae (PPISNe) have long been invoked to explain
the $m_1 \sim 35\, \Msun$ feature as a pile-up of masses at the instability
onset~\citep[see e.g.][]{Woosley:2016hmi,Talbot:2018cva}. 
Recent studies, however, have shown this hypothesis to be in conflict 
with theoretical predictions of the expected location of the related mass gap 
or the relative number of systems in this mass 
range \citep{2019ApJ...882..121S,Farag:2022jcc,Hendriks:2023yrw}.
Moreover, the PPISN channel would require extreme values of nuclear reaction rates,
in particular the $^{12}\mathrm{C}(\alpha,\gamma)^{16}\mathrm{O}$ reaction rate,
to match the BH mass distribution inferred from GW 
observations~\citep[see e.g.][for recent analyses]{Tong:2025wpz,Antonini:2025ilj}.
Also, while the interplay between time delays and the 
dependence of BH maximum mass  on progenitor metallicity can give rise to an 
evolution with redshift of the BH mass spectrum~\citep{2019ApJ...887...53F,Mukherjee:2021rtw}, 
this scenario relies preferentially on longer time delays, which are not favored by our analysis.
The results presented in this work add to the mounting evidence that
this formation channel may not be the preferred one to explain
the $m_1 \sim 35\, \Msun$ feature in the GW spectrum. 

Peaks in the BBH mass spectrum at $m_1 \gtrsim 30 \, \Msun$ can also be caused by low-metallicity stars 
whose structure has not been altered by binary interaction~\citep{Schneider:2020vvh,Schneider:2023mxe}, 
which then get paired through other processes. The low-metallicity requirement naturally ties this process
to redshift through the chemical enrichment of the Universe, but further work is required to assess
the compatibility of these models with our results. 

We also recover a low-mass peak \mbox{$\Mtot \sim 20 \, \Msun$}  \mbox{$(m_1\sim10\,\Msun)$}
at $z = 0.2$ (current uncertainties prevent confirmation at $z = 1$). 
This previously reported feature~\citep{LIGOScientific:2020kqk,KAGRA:2021duu,LIGOScientific:2025pvj} 
is well-explained by isolated binary evolution at near-solar
metallicity~\citep{Fragos:2022jik,Agrawal:2023ecm}
or dynamical evolution of stars and compact objects in metal-rich dense clusters~\citep{Ye:2025ano}.
More observations are needed to confirm the invariance of this low-mass peak across redshift.

\section{Conclusions}

We presented a non-parametric reconstruction of the mass-redshift correlation 
in merging BBH using Delaunay triangulation and barycentric interpolation.
Our analysis of GWTC-4.0 data reveals that the high-mass feature at $\Mtot \sim 70\,\Msun$
appears predominantly from $z \gtrsim 0.7$ and is absent at $z = 0.2$. 
At the same time, we caution against overinterpreting (the medians of)
non-parametric reconstructions, which often, including here, come with large statistical errors.

So, where would the heavy binaries have gone?
Taken together, our results suggest the presence of a population with \mbox{$\Mtot \sim 70\Msun$}, 
characterized by a fast-merging component that dominates at high redshift
and a slow-merging component that becomes increasingly important at low redshift.
While such a trend could arise from multiple formation pathways---such as dense environments for
short delays and isolated (and ejected) binaries for long delays---establishing the relative 
contribution of specific channels requires further observational and theoretical work. 
This includes investigating potential correlations in other parameters, most notably spins and eccentricity.
 
Moreover, our findings imply that analyses measuring cosmological parameters from 
the redshift dependence of the detector-frame mass 
spectrum~\citep{Chernoff:1993th,Markovic:1993cr,Farr:2019twy,Mastrogiovanni:2021wsd,LIGOScientific:2025jau}
should carefully account for the evolution of the prominence of the high-mass peak with $z$.

The key strength of using Delaunay triangulation is that the interpolation skeleton,
including the number and location of the nodes, is inferred directly from the data.
In practice, this optimized complexity reduces the number of parameters by one to two orders 
of magnitude compared to the pixelized approach by \cite{Heinzel:2024hva}, and attain a good
recovery of arbitrary functional forms at a lower tuning cost compared to the binned
Gaussian-process approach by \cite{Ray:2023upk}, both of which rely on a fixed grid.
For this reason, we expect our flexible framework to be
significantly more efficient at probing correlations in dimensions $\geq 3$,
making it ideally suited for the forthcoming big-data era of GW astronomy.


\begin{acknowledgements}

We thank April Qiu Cheng, Raffi Enficiaud, Jack Heinzel, and Matthew Mould for discussions.
R.T., A.T., T.B. and D.G. are supported by
ERC Starting Grant No.~945155--GWmining, 
Cariplo Foundation Grant No.~2021-0555, 
MUR PRIN Grant No.~2022-Z9X4XS, 
Italian-French University (UIF/UFI) Grant No.~2025-C3-386,
MUR Grant ``Progetto Dipartimenti di Eccellenza 2023-2027'' (BiCoQ),
and the ICSC National Research Centre funded by NextGenerationEU. 
A.T.~and D.G.~are supported by MUR Young Researchers Grant No. SOE2024-0000125.
D.G.~is supported by MSCA Fellowship No.~101149270--ProtoBH.
Computational work was performed 
at CINECA with allocations through INFN and the University of Milano-Bicocca, 
and at NVIDIA with allocations through the Academic Grant program.
This research has made use of data or software obtained from the Gravitational Wave Open Science Center. 
\end{acknowledgements}

\software{
    astropy~\citep{Astropy:2013muo}, 
    eryn~\citep{Karnesis:2023ras},
    numpy~\citep{Harris:2020xlr}, 
    qhull~\citep{Barber:1996lmi},
    scipy~\citep{Virtanen:2019joe}.
}
\bibliography{triangulation}
\bibliographystyle{yahapj_edited}

\appendix

\section{Delaunay triangulation for inference}\label{app:delaunay}

We seek to represent general multidimensional distributions using an agnostic model with a low computational cost.
Our strategy is to construct a trans-dimensional interpolator for the merger rate using barycentric
interpolation on a Delaunay triangulation~\citep{Delaunay_1934aa, RIPPA1990489, rajan1994}.
The number of triangulation points is allowed to vary through inference,
thus modeling arbitrarily simple (or complex) distributions depending on the available information in the data. This method can be applied to parameter spaces of arbitrary dimensionality. While we restrict our application of the method to a two-dimensional parameter space, we describe it in full generality.

We model the differential number of events $\rate(\theta;\Lambda)$ in dimension $D$. 
The hyperparameters $\Lambda$ we aim to infer
are the number of triangulation vertices $\Nv$, their locations $v_k$, and their weights $w_k$,
where the $v_k$'s live in the same space as $\theta$ and the $w_k$'s are real numbers.

In our current implementation, we infer the rate within a $D$-dimensional box $\mathcal{D}$  
so that $\rate(\theta) = 0$ if $\theta \not\in \mathcal{D}$. To do so, we place $\Nc = 2^D$ fixed vertices
at the corners of the domain; their associated weights $\many{W_k, k=1, \dots, \Nc}$ are allowed to vary.
Thus,
\begin{equation}
    \Lambda = \many{\Nv, \many{W_k}_{k=1, \dots, \Nc}, \many{v_k, w_k}_{k=1 \dots, \Nv - \Nc}} \, .
\end{equation}
The number of free parameters is
\begin{equation}
    \Np = 1 + \Nc + (D + 1) (\Nv - \Nc) 
    = 1 - D \,2^D+  (D +1) \Nv  \,.
\end{equation}

We denote the (natural log) differential rate as computed by a triangulation with
parameters $\Lambda$ by \mbox{$\log\rate(\theta | \Lambda) \equiv \tri(\theta|\Lambda)$}.
To compute the differential rate at a given point $\theta$, we find the simplex $S(\theta)$ (i.e. triangle if $D=2$)
wherein it lies, compute its barycentric coordinates, and use them to linearly interpolate the weights at their vertices:
\begin{equation}
	\tri(\theta| \Lambda) = \sum_{v \in S(\theta)} w_v b_v(\theta) \,.
\end{equation}
We illustrate this process for $D=2$ in Fig.~\ref{fig:tri_tikz}. In such case, each triangle $S(\theta)$
has three pairs of vertices and weights $\many{(v_1, w_1), (v_2, w_2), (v_3, w_3)}$ from which one can
construct three barycentric coordinates $\{b_1(\theta), b_2(\theta), b_3(\theta)\}$.

The number of expected events within a domain $\mathcal{D}$ for a given triangulation $\Lambda$ is given by 
\begin{equation}
	N(\Lambda;\mathcal{D}) = \int_{\theta \in \mathcal{D}} \mathrm{d}\theta \, e^{\tri(\theta;\Lambda)}\,
    \label{eq:N}\,.
\end{equation}
For a single simplex $S$, Eq.~\eqref{eq:N} can be expressed in closed form as
\begin{equation}
    N(\Lambda;S) = D! \, |S|  \, \sum_{i=1}^{D+1} \frac{\exp(w_i)}{\prod_{j\neq i} (w_i - w_j)}\,,
\end{equation}
where $|S|$ denotes the generalized volume of $S$ (i.e. the area if $D=2$) and $w_{1}, \dots, w_{D+1}$ 
are the weights of its vertices. 

We construct Delaunay triangulations using \texttt{qhull}~\citep{Barber:1996lmi} via \texttt{scipy}~\citep{Virtanen:2019joe}.

\section{Mass-ratio distribution\label{app:q}}

\begin{figure}
    \centering
    \includegraphics[width=0.49\linewidth]{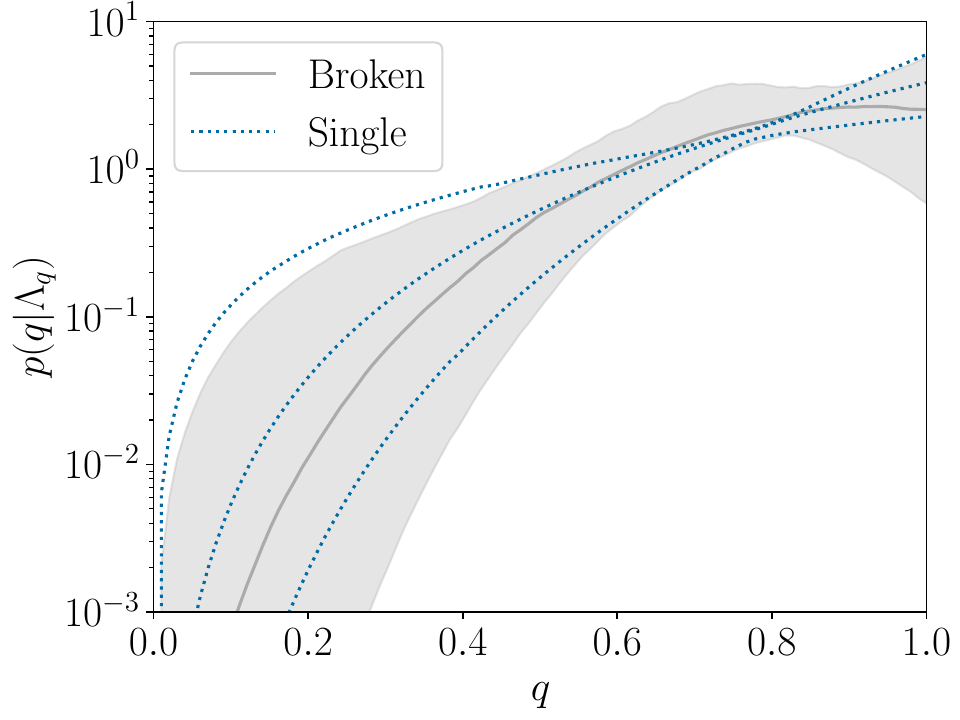}
    \includegraphics[width=0.49\linewidth]{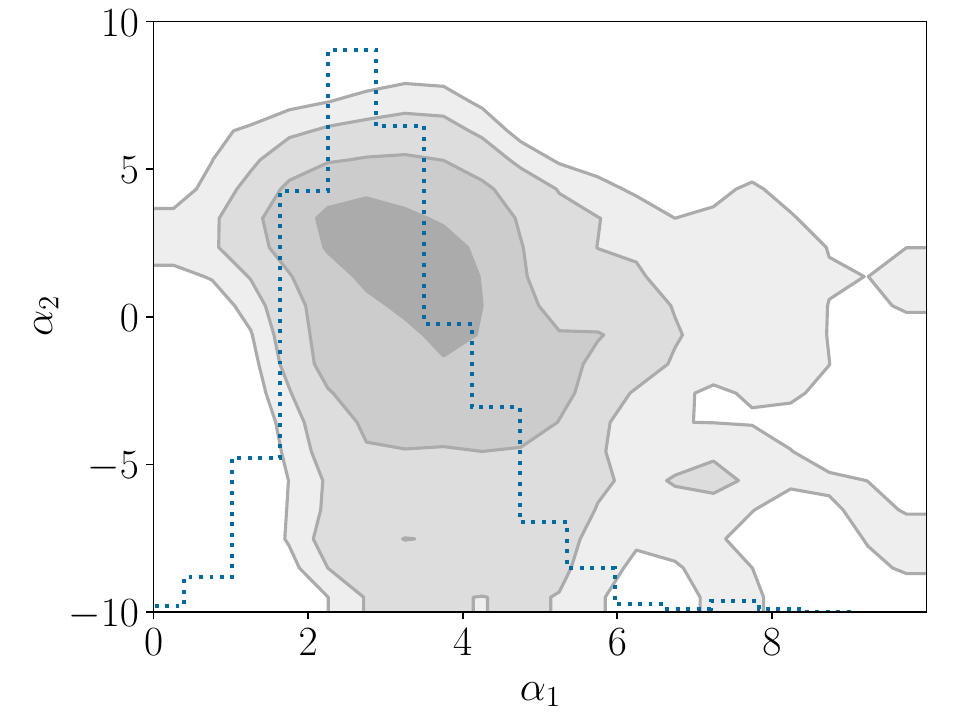}
    \caption{
    \emph{Left:} Posterior predictive distribution of the mass-ratio distribution.
    The red (grey) solid line represent the median, while the shade (dotted) envelope
    contains the 90\% credible interval assuming a broken (single) power-law distribution.
    \emph{Right:} Posterior distribution of the power-law indices. The gray area represents
    the posterior probability on $(\alpha_1, \alpha_2)$ for the broken power-law model, while
    the dotted histogram represents the posterior probability on $\alpha_1$ for the single
    power-law model.
    }
    \label{fig:all_q}
\end{figure}

The fiducial model by~\cite{LIGOScientific:2025pvj} describes the joint $(m_1, q)$ distribution as a sum of components, 
each independent in $m_1$ and $z$. This formulation makes the overall distribution non-separable in $m_1$ and $q$, 
and this cannot be straightforwardly implemented in our framework.
Guided by \cite{LIGOScientific:2025pvj}'s findings (see their Fig. 5), we describe the mass-ratio distribution using a broken power law:
\begin{equation}
    p(q|\Lambda_q) \propto \begin{cases} 
    0 & q < q_{\mathrm{min}}\,, \\
    (q / q_{\mathrm{cut}})^{\alpha_1} & q_{\mathrm{min}} < q < q_{\mathrm{cut}}\,,\\
    (q / q_{\mathrm{cut}})^{\alpha_2} &  q_{\mathrm{cut}} \leq q\,.
    \end{cases} \, .
\end{equation}
Prior distributions for $\Lambda_q = (q_{\mathrm{cut}}, \alpha_1, \alpha_2)$ 
are uniform across $(0.5, 1.0)$, $(0, 10)$, and $(-10, 10)$ respectively. 
The lower end $q_{\mathrm{min}} = 3 \, \Msun / m_1$ is chosen so that $m_2 > 3 \, \Msun$.
The population predictive distribution of $q$ and the hyperposterior distribution of 
$(\alpha_1, \alpha_2)$ are shown in Fig.~\ref{fig:all_q}. 
We observe an enhanced number of mergers at $0.6 \lesssim q \lesssim 0.8$ compared
to the single power-law model. 

For completeness, we also show the results for a single power law with,
which corresponds to $q_{\mathrm{cut}} \rightarrow 1$ and/or $\alpha_2\to \alpha_1$. 
The conclusions of our work regarding the behavior of the merger rate
across $(M_{\mathrm{tot}}, z)$ are overall unchanged.
The main quantitative difference is that the credibility for an enhanced rate at
$\sim 70 \, \Msun$ is lower ($\sim 94\%$ vs. $\sim 98\%$).

\section{Formalism for hierarchical inference}\label{sec:selection}

We express the population likelihood for observing $N_\mathrm{obs}$ events with data \mbox{$\{d_j\} = \{d_1,\dots,d_{N_\mathrm{obs}}\}$} for a set of population hyperparameters $\Lambda$ as~\citep{Mandel:2018mve,Vitale:2020aaz}
\begin{equation}
    p(\{d_j\} | \Lambda) \propto 
    e^{-\Ndet(\Lambda)} \prod_{j=1}^{N_{\mathrm{obs}}} 
    \left\langle \frac{\rate(\theta | \Lambda)}{\pi_{\mathrm{PE}}(\theta)} \right\rangle_{\theta \sim p(\theta | d_j)}\,,\label{eq:pop_logl}
\end{equation}
where $p(\theta | d_j)$ and $\pi_{\mathrm{PE}}(\theta)$ are the posterior and prior distribution of the parameters describing event $j$. 
The number of expected events $N_{\rm det}(\Lambda)$ is given by \begin{equation}
N_{\rm det}(\Lambda) = \left\langle 
\frac{\rate(\theta | \Lambda)}{\pi_{\mathrm{inj}}(\theta)} p(\mathrm{det}| \theta)
\right \rangle_{\theta \sim \pi_{\mathrm{inj}}(\theta)}
\label{eq:selection}
\end{equation}
where 
\begin{equation}
    p({\rm det}|\theta) = \int_{d > \text{threshold}}   p(d|\theta)\, {\rm d}d , \label{eq:pdet}
\end{equation}
is the probability of detecting a source with parameters $\theta$, $p(d|\theta)$ is the likelihood for data $d$, and the integral is restricted to realizations $d$ that exceed the detection threshold (defined via the chosen ranking statistic).

For cosmological calculations, we assume the 
\texttt{Planck15} cosmology as implemented in \texttt{astropy}~\citep{Astropy:2013muo}. 

Equations~\eqref{eq:pop_logl} and~\eqref{eq:selection} are computed via Monte Carlo integration, here indicated with $\langle \cdot \rangle$. Selection effects are estimated using the injection campaign presented by~\cite{Essick:2025zed}.
We set the population likelihood to zero for hyperparameters
that yield a variance in the Monte Carlo estimators greater than one \citep{Talbot:2023pex,LIGOScientific:2025pvj,Mancarella:2025uat}.
We verify the robustness of this choice by re-running our inference using thresholds of 0.8 and 1.2.
The results are qualitatively unchanged; the credibility for an enhanced merger rate at
$\sim 70 \, \Msun$ is $\sim 95\%$ instead of $\sim 98\%$ for the 1.2 threshold.

As discussed in the main text, the prior on the modeled part of the population parameters
$(\Lambda_q, \Lambda_\chi, \Lambda_{\vartheta})$ are taken equal to those of~\citet{LIGOScientific:2025pvj},
while the prior on the triangulation parameters is uniform across the area of interest 
for the triangulation vertices and along $(-20, 15)$ for the triangulation weights. 
The implied prior range on $\rate$ (or equivalently $\volrate$) is compatible with
 (if not more conservative than) those used in similar non-parametric approaches~\citep{Heinzel:2024hva,Heinzel:2024jlc, Alvarez-Lopez:2025ltt}.
We also verified the robustness of our results against prior choices
by extending the  weight prior down to $(-25, 15)$, which lowers the induced prior on $\rate$ 
by two to three orders of magnitude. 
The results for $z = 0.2$ remain unchanged. For $z=1$, the lower end of the posterior 
distribution on $\volrate$ lowers so that the credibility for 
$\volrate(z = 1) > \volrate(z = 0.2)$ is slightly lower ($\sim 94\%$ instead of $\sim 98\%$). We stress that the key result of this work, namely the disappearance of the peak at low redshifts, is not affected by prior choices.
The lack of an obvious choice for the prior range of $w_i$ suggests non-parametric population models 
are better suited to conduct inference in the observed space, 
where the domain of inference is naturally limited to data-dominated regions~\citep{Toubiana:2025syw}.

We also compute the Kullback-Leibler divergence $D_{\mathrm{KL}}$~\citep{kl_div}
to compare the posterior volumetric merger rate to its prior distribution at any given $(\Mtot, z)$. 
The prior distribution on $\log_{10} \volrate(\Mtot, z)$, denoted by $p(\log_{10} \volrate)$ for simplicity, 
is constructed by sampling  the prior distribution of $\Lambda_{\tri}$ 
(i.e. number of vertices, positions, and weights) and computing the corresponding 
$\log_{10} \volrate(\Mtot, z)$. The posterior distribution, denoted $p(\log_{10} \volrate | \{d_j\})$, 
is computed analogously by sampling from $p(\Lambda_{\tri} | \{d_j\})$. To approximate these probability
densities, we generate $10^4$ samples in every $(\Mtot, z)$ of interest and fit a 
Gaussian Kernel Density Estimator. The Kullback-Leibler divergence (in nat) for any given 
$(\Mtot, z)$  is then computed as
\begin{equation}
    D_{\mathrm{KL}} = \left\langle 
    \ln\frac{p(\log_{10} \volrate | \{d_j\})}{p(\log_{10} \volrate)}
    \right\rangle_{\log_{10} \volrate \sim p(\log_{10} \volrate | \{d_j\}) } \,.
\end{equation}  

\section{Population inference for $\protect\lowercase{(m_1, z)}$\label{app:m1_z}}

\begin{figure*}
    \centering
    \includegraphics[width=0.49\columnwidth]{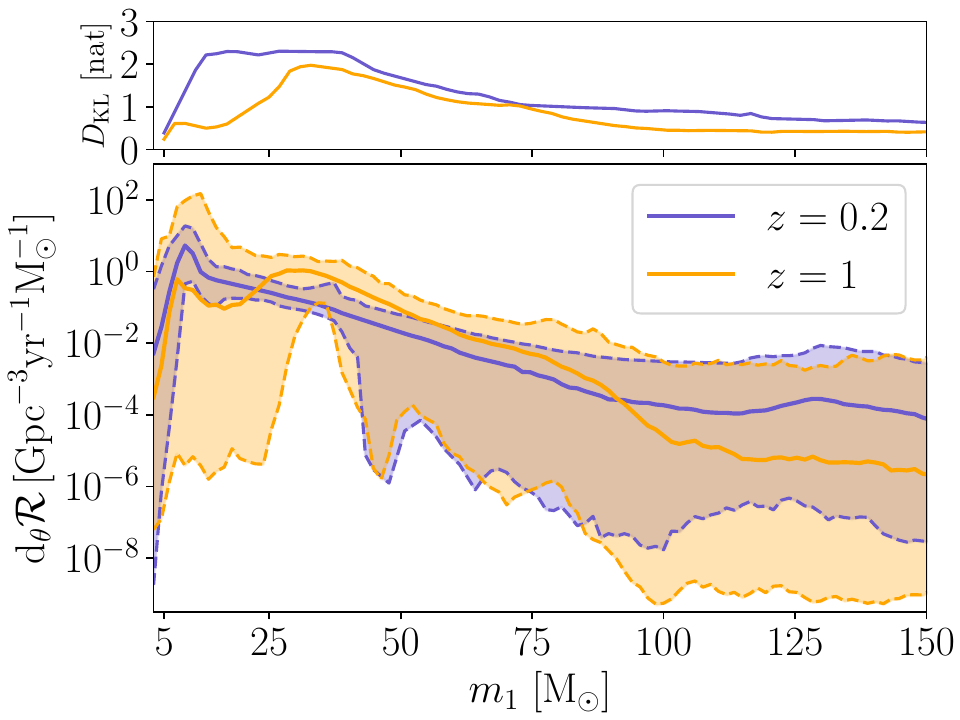}
    \includegraphics[width=0.49\columnwidth]{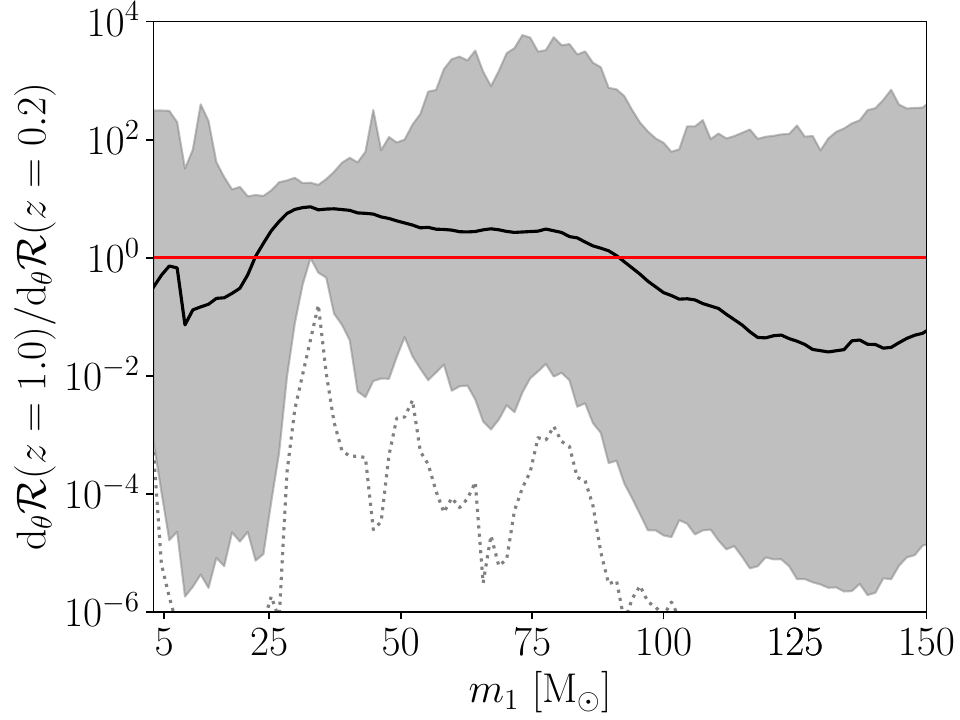}
    \caption{
        Joint reconstruction of the $(m_1,z)$ BBH volumetric merger rate, 
        mirroring Fig.~\ref{fig:Mtot_z} in the main body of the paper. A constant rate is excluded with 95\%
        credibility (i.e. the solid red line is included in the 90\% symmetric credible interval, 
        here shaded in grey).
    }
    \label{fig:m1_z}
\end{figure*}

To facilitate comparison with other studies, Fig.~\ref{fig:m1_z} shows the results of our method applied to the joint $(m_1, z)$ distribution instead of $(\Mtot, z)$. The significance of the enhanced rate at $\sim 35 \,\Msun$ at $z=1$ versus $z=0.2$ is slightly lower than for $(\Mtot, z)$
($\sim 95\%$ vs.$\sim 98\% $, right panel of Fig.~\ref{fig:m1_z}),
likely due to larger measurement uncertainties in $m_1$, 
but the qualitative disappearance of the high-mass peak at low redshift persists. 

\end{document}